# The Pareto-frontier-based Stiffness of A Controller: Trade-off between Trajectory Plan and Controller Design

Zhe Shen and Takeshi Tsuchiya
Department of Aeronautics and Astronautics, The University of Tokyo, Japan
Contact: Zhe Shen, zheshen@g.ecc.u-tokyo.ac.jp

**Abstract**

Approaching a set goal for a UAV comprises a trajectory plan and a controller design (control after plan problems). The optimal trajectory (reference) is calculated before being tracked with a proper controller. It is believed that the quadrotor will follow the designed trajectory totally in the trajectory plan process. However, the dynamic state error usually, for a mismatched feed-forward, spoils this assumption, making the unwanted sacrifice in the objective function defined in the trajectory plan process. We base the target problem in this research on a second-order system model which widely exists in vehicles' dynamics. Specially, the unavoidable dynamic state error is considered in the trajectory plan process, assuming the LQR without the feed-forward is applied in the subsequent control after plan problems. The Copenhagen Limit provides the possibility of estimating the dynamic state error in an analytical solution. The trade-off results are provided in multiobjective Pareto front solutions and the mapped 'pseudo' Pareto fronts. We explore the relationship between the controller and the corresponding 'pseudo' Pareto fronts.

## I. INTRODUCTION

Higher demands on UAV flying tasks inspire research in solving the multiobjective optimization trajectory design problems. Typical objectives comprise least fuel consumption [1], shortest flying time [2], safety concern [3], etc. The number of the objectives in multiobjective trajectory/path optimization can be double [2,4], triple [5,6], or even more [7].

A typical problem in the multiple-objective control problems with the analytical solution is the trade-off between dynamic state error and the control effort [8].

$$J = \int_0^\infty [x(t)^T R_1 x(t) + u(t)^T R_2 u(t)] dt$$

Here, the dynamic state error can be analytically presented since the reference is a setpoint. Together with the infinite time horizon, the analytical solution exists. Unfortunately, only numerical results can be received even with a slight change in this problem [9].

The dynamic state error loses the analytical form for a moving reference in a general control problem without using a feed-forward signal. Here, the reason for prohibiting the feed-forward signal in this research is that several numerical solvers introduce the desired feed-forward signal's mismatch with the desired trajectory. Typical research endeavoring to decrease this mismatch can be found in [10,11].

Since utilizing this feed-forward introduces the dynamic state error, we are to discard it in a trajectory design problem result. It means that the only information used in the trajectory design process is the time-specified trajectory itself. Thus, a proper formula to calculate/estimate the resultant dynamic state error in a tracking control problem is demanded.

[12] proposed an analytical solution of the dynamic state error in a limit form. This limit estimates the dynamic state error of some particular controlled (LQR) systems. And we consider the estimation of the dynamic state error in trajectory design based on [12] in this research.

We solve the multiobjective optimization problem in this research using the weighted sum method [13]. This method is widely used in generating the resultant Pareto frontier. And a variety of researches pick the compromised optimal result near the elbow region, a region universally exists in a Pareto frontier [14]. Besides the conventional Pareto frontier, we also mapped the corresponding tracking control result as a 'pseudo Pareto frontier' to better analyze the relationship between the controller and the trajectory. The optimum solver is direct collocation (trapezoid) since this method has high compatibility with the limit form in dynamic state estimation.

The tracking control result in 'pseudo Pareto frontier' shows the atypical property. The similar atypical property can also be found in Schaffer's function [15]. However, the relevant Pareto frontier in [15] is not empowered with a physical model; the atypical property rarely happens in a natural system.

We model this result with a physical model, a Hooke's law applicable spring, and analyze its relationship with the set controller parameters.

This paper is organized into seven sections. In Section II, the considered second-order system and the task allocation are illustrated. Section III explains the method of the trajectory trade-off scheme. And the relevant simulation setup is detailed in Section IV. Its results are presented and modeled in Section IV and Section VI, respectively. Finally, we make conclusions and further discussions in Section VII.

## II. MODEL AND TASK ALLOCATION

***System Model***

The model analyzed in this research is a general second-order system. This model notes its importance in a variety of mechanical systems, e.g., UAV altitude controller design is one of them. The reason is that the generalized second-order system model has a parallel in the corresponding Newtown's law.

Without losing generality, a planar quadrotor is modeled to parameterize a second-order system model (Figure 1). This empowers the result a practical physical meaning in this research.

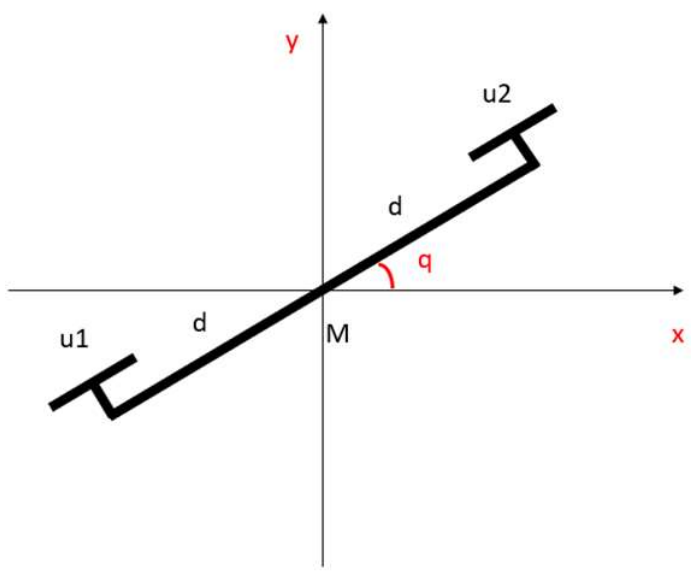

Fig. 1. The planar drone model is used in this research.

A planar quadrotor is a 2-D quadrotor where the vehicle is restricted to move within a vertical plane. The gravity acts in the negative $y$-direction. The air resistance and noise are neglected in this research to avoid the unexpected bias in our numerical verification. Its parameters are unanimous with [12] and are listed in Equation (1) – (3).

$$M = 0.54\ Kg \tag{1}$$

$$d = 0.12164\ m \tag{2}$$

$$g = 9.81\ m/s^2 \tag{3}$$

$M$ represents the quadrotor's mass which is totally contributed by two rotors. And $d$ represents the length of each quadrotor's arm. $g$ is the gravitational acceleration.

In the following, we use $u_1$ and $u_2$ to represent the thrusts provided by the left rotor and right rotor, respectively. The saturation of the thrust is not considered in our mathematical verification.

Besides, the position vector of the quadrotor's geometric center, $[x\ y]$, is stated with respect to the earth frame. And its attitude is described using the angle, $q$, rotating from the earth frame to the UAV's body-fixed frame. Based on these, the state vector is defined in Equation (4).

$$r = \begin{bmatrix} x \\ y \\ q \end{bmatrix} \tag{4}$$

The dynamics are deduced based on Newton–Euler equations, Equation (5)–(6).

$$F = m \cdot a \tag{5}$$

$$\tau = I \cdot \alpha \tag{6}$$

The resulting dynamic equations are Equation (7) – (9).

$$\ddot{x} = -\frac{1}{M} \cdot \sin(q) \cdot (u_1 + u_2) \tag{7}$$

$$\ddot{y} = \frac{1}{M} \cdot (u1 + u2) \cdot cos(q) - g \tag{8}$$

$$\ddot{q} = \frac{1}{M \cdot d} \cdot (u2 - u1) \tag{9}$$

A variety of researches design the controller with a linearized UAV model in the hovering state. Typically, [16] analyzes the region where the linearization-based controller guarantees stability. We apply the same method here. After linearization, Equation (8) becomes a general linear second-order system, Equation (10).

$$\ddot{y} = \frac{1}{M} \cdot (u1 + u2) - g \tag{10}$$

This research uses Equation (10) as our general second-order system model for later experiments.

***TASK ALLOCATION***

The quadrotor is to experience a vertical take-off (VTO) with some requirements.

The typical method in such tasks is a control-after-plan problem; a trajectory is designed first before it is tracked using a 'well-designed' controller. A well-designed controller usually contains the feed-forward signal from the trajectory design part, since a well-defined feed-forward signal helps the control process. However, for the reasons mentioned beforehand, the feed-forward signal is prohibited here. Instead, a linear quadratic regulator (LQR) without feed-forward is used in the altitude tracking process.

Firstly, a quadrotor's trajectory is designed with the following requirements:

1. The initial altitude is 0, the initial velocity is 0, the initial acceleration is 0.

2. The final altitude is 5 meters, while the final velocity and the final acceleration are not restricted.

3. The total task is completed within 1 second.

4. The following objective function, Equation (11), is evaluated and is expected to be minimized.

$$\min \int_0^1 (\ddot{x}^2 + \ddot{y}^2 + \ddot{q}^2) dt \tag{11}$$

5. Since we have decided the later LQR used in the tracking process, the dynamic state error can be estimated in the trajectory plan process. It is detailed in Section III.

Secondly, a controller (LQR) is designed to track the designed trajectory. The tracking LQR is designed with the following requirements:

1. There is no steady-state error. It means that the quadrotor is to reach the desired setpoint if sufficient time is given. However, the actual final position at 1 second can be a position other than the desired setpoint since the control process ends at 1 second rather than infinite.

2. The desired time-specified trajectory itself is the only information allowed to be used as the reference. The feed-forward signal is not allowed in the altitude control.

3. The parameters of the controller are designed before the Trajectory Design process.

Notice that the actual moving trajectory can never collapse the desired trajectory designed in the Trajectory Design process without the feed-forward signal; the dynamic state error is unavoidable. The dynamic state error is the source of the control signal.

Our ultimate goal is to find the actual moving trajectory with the least cost function, Equation (11). Note the acceleration used to calculate the actual cost function is the actual acceleration while tracking.

## III. CONTROLLER AND TRAJECTORY TRADE-OFF SCHEME

***Altitude Controller***

The altitude controller used in tracking control is an LQR.

The altitude LQR utilized in this research is similar to the altitude LQR in [12]. It is detailed in Equation (12)-(13).

$$\begin{bmatrix} \dot{y} \\ \ddot{y} \end{bmatrix} = \begin{bmatrix} 0 & 1 \\ 0 & 0 \end{bmatrix} \begin{bmatrix} y \\ \dot{y} \end{bmatrix} + \begin{bmatrix} 0 \\ \frac{1}{M} \end{bmatrix} (u_1 + u_2) + \begin{bmatrix} 0 \\ -g \end{bmatrix} \tag{12}$$

$$u_1 + u_2 = -[k_1 \;\; k_2]\begin{bmatrix} y \\ \dot{y} \end{bmatrix} + [N_1 \;\; N_2]\begin{bmatrix} reference \\ 0 \end{bmatrix} + Mg \quad (13)$$

$k_1$ and $k_2$ are the control constants; $N_1$ and $N_2$ are the offset constants; $reference$ is the position reference generated from the Trajectory Design process.

**Theorem:** For a stable LQR-controlled system, if $k_1$, $k_2$, $N_1$, and $N_2$ in Equation (13) satisfies:

$$k_1 = N_1 \quad (14)$$

Then the steady-state error is 0.

**Proof:** see [12].

Once the eigenvalue pairs are determined, we design the LQR based on Equation (14) to guarantee zero steady-state error.

In our experiments, the eigenvalue pairs in each of four experiments are [-10 -100], [-20 -200], [-30 -300], [-50 -500], respectively. Note that they are determined before the trajectory generation process.

***Attitude and Position Controller***

The attitude and position controllers are two PID controllers with feed-forward. Note that the feed-forward is not allowed in the altitude controller; this research presumes that the feed-forward is not used in tracking control. In contrast, the UAV's attitude and position controls are encouraged to use the feed-forward information since we do not expect the dynamic state error in attitude. The whole movement of the UAV should be in the 1-D linear movement to support our mathematical analysis.

Equation (15)-(16) gives the relevant settings in our PID controllers.

$$\ddot{x} = \ddot{r}_x + 10 \cdot (\dot{r}_x - \dot{x}) + 100 \cdot (r_x - x) \quad (15)$$

$$\ddot{q} = \ddot{r}_q + 80 \cdot (\dot{r}_q - \dot{q}) + 100 \cdot (r_q - q) \quad (16)$$

***Trajectory Design***

The difference between a path and a trajectory is that a trajectory is a time-specified function [17]. Although the path has been already defined-- a 5-meter vertical path in 1 second-- in TASK ALLOCATION, it is not specified with time. Undeniably, the trajectory can be directly designed as a constant velocity one. However, a better solution, e.g., lower cost function in Equation (11), may be found by assigning the time to the path properly.

Our ultimate goal is to receive the lowest cost function in Equation (11) in our actual tracking result.

The conventional way to solve this problem is to design the following optimal trajectory design problem:

$$\min \int_0^1 \ddot{y}^2 dt \quad (17)$$

Subject to

$$y(t=0) = 0 \quad (18)$$

$$\dot{y}(t=0) = 0 \quad (19)$$

$$y(t=1) = 5 \quad (20)$$

$$\ddot{y} = \frac{1}{M} \cdot u - g \quad (21)$$

$$0 \leqslant y \leqslant 5 \quad (22)$$

The analytical solution does not exist for a general optimal trajectory design problem. To solve this problem, a variety of numerical methods are developed.

It is worth mentioning that tracking this resulting trajectory using the designed controller, Equation (12)–(13), will introduce the dynamic state error. The dynamic state error is the source of the feedback.

It is undeniable that the dynamic state error is acceptable for our tracking result, since the object will follow our planned trajectory in general. However, the existence of the dynamic state error indicates that the UAV does not follow the designed time-specified trajectory totally. True, unless the dynamic state error is zero, e.g., the actual trajectory overlaps the desired trajectory, the UAV will not fully follow the designed trajectory.

Consequently, the actual movement of the UAV will not meet the optimal solution to Equation (17)–(22). Although the designed trajectory solves this problem, the UAV does not fully follow it. In fact, fully following it is impossible without a feed-forward, as we restricted, since the dynamic state error is the feedback-control signal's source.

It can take several potential measures to decrease this dynamic state error. The way utilized in this research is to consider the dynamic state error while defining the objective function, Equation (17).

Instead of using the conventional objective function defined in Equation (17), we consider a modified objective function, Equation (23).

$$\min \int_0^1 \left(\ddot{y}^2 + \mu \cdot e(t)^2\right) dt \quad (23)$$

Here, $e(t)$ represents the dynamic state error estimation, e.g., Equation (24)-(25), which has been thoroughly discussed in [12].

$$e(t,n) = \frac{t}{n} \sum_{i=1}^{n} \left[ v_{ref}\left(\frac{t}{n} i\right) \cdot (1-p)^{n+1-i} \right] \quad (24)$$

$$e(t) = \lim_{n \to \infty} e(t,n) \quad (25)$$

We utilize Equation (25) to estimate the dynamic state error in Equation (23) since both the system and the controller are continuous. In addition, the eigenvalue pairs in the tracking control, [-10 -100], [-20 -200], [-30 -300], [-50 -500], are degraded to the single eigenvalues, -10, -20, -30, -50, respectively, to accommodate the dynamic state error estimation in Equation (24)-(25).

The coefficient, $\mu$ in Equation (23), is the weight for the dynamic state error. Equation (23) provides the trade-off scheme between the trajectory design and the controller ability consideration, dynamic state error in specific.

Putting a significant weight on the dynamic state error tends to receive a less dynamic state error in the tracking result. However, the adverse effect is that the optimality might be affected. Here, it means that the resulting trajectory planed will not be the one with the lowest acceleration. The larger the weight is, the more compromise is made in optimality.

Vice versa.

Note that the above trade-off scheme happens in the trajectory plan process only; a particular trajectory is designed for each dynamic state error weight with the pre-defined controller. In comparison, the corresponding resulting actual trajectory in the tracking process may not inherit a similar trade-off unless the actual tracking trajectory overlaps the designed trajectory for being tracked. For this reason, the actual dynamic state error and the actual acceleration might not necessarily be compromised in a way happening in the trajectory design process.

This research analyzes the mentioned trade-off schemes between the dynamic state error and the objective – the acceleration—both in the trajectory design process and the tracking control result.

***Experiments and Testbed***

In general, each experiment set consists of two sections: the optimal trajectory design part and the tracking control of the designed trajectory part.

Different from conventional researches, the controller is predefined before the optimal trajectory design process. There are four experiments in total. We tune the altitude controllers with different eigenvalue pairs, [-10 -100], [-20 -200], [-30 -300], [-50 -500], respectively, in each experiment. We will take [-20 -200] as a demo to further explain our experimental procedures.

The first step is to decide the dynamic state error function, Equation (24)–(25). Calculation in this step requires a first-order eigenvalue. The equivalent dominant eigenvalue for the eigenvalue pair [-20 -100] is -20.

Once the dynamic state error function is obtained, the next step is to solve a multiple-objective optimization problem. The trade-off scheme's objectives, Equation (23), are the acceleration and the dynamic state error.

We use the weighted sum method [13] to transfer the multiobjective optimization problem into a single-objective optimization problem with a varying coefficient, $\mu$ in Equation (27). The constraints are in Equation (18)–(22).

The method for solving this optimization problem is direct collocation (trapezoid); the first step in the direct method is to discretize. We decide to use the direct method since the dynamic state error estimator, Equation (24)–(25), also possesses a discrete nature.

The platform to solve the trajectory design task is the script, MATLAB. A further explanation of the relevant program is in IV SOLVER SETUP. For a deep understanding of this solver, [18] is recommended.

A Pareto frontier for trajectory design is received after repeating the optimal trajectory design problem with different dynamic state weights. This Pareto frontier in the trajectory design process demonstrates the trade-off between the predicted dynamic state error and the cost (designed acceleration).

On the other hand, each trajectory design problem with a unique dynamic state error weight in Equation (23) outputs a resulting trajectory. This resulting trajectory is further used as the reference in the subsequent tracking control.

We simulate the tracking control in a 2-D UAV simulator, Figure 3, written in SIMULINK, MATLAB. It comprises two sections, the Controller Section and the Dynamics Section.

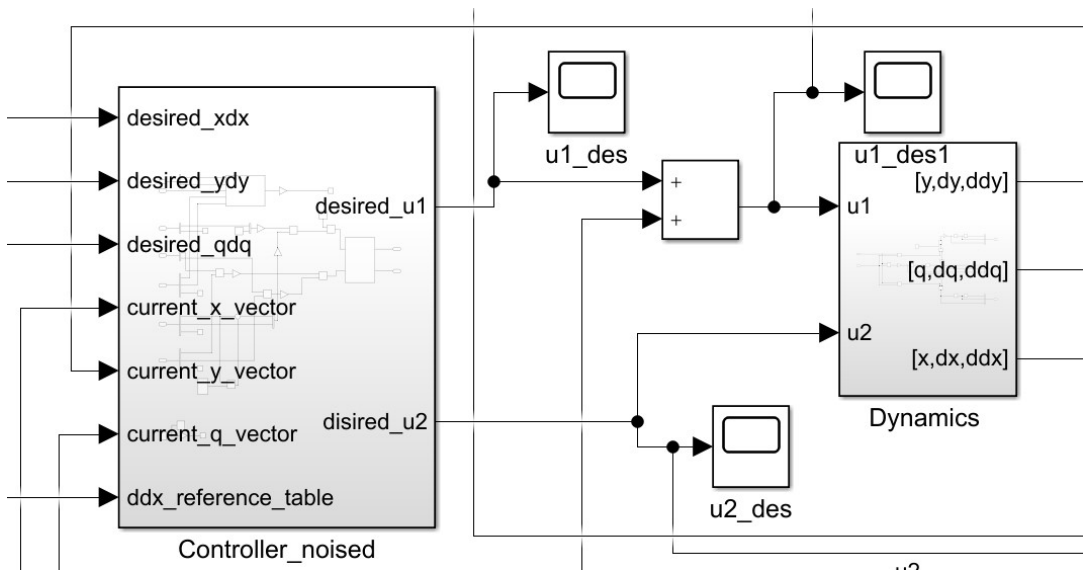

Fig. 3. The dynamic system simulator is written in SIMULINK.

We equip the altitude controller and the attitude controller in the Controller Section in SIMULINK. The altitude controller with the eigenvalue pair, [-20 -200], is specified in Equation (12)–(14). Furthermore, the attitude controller is set based on Equation (15)–(16).

The Dynamics Section in the simulator is based on the UAV's dynamic equations, Equation (7)–(9).

This simulator's output is the tracking result (actual trajectory in tracking) once we input a designed trajectory (reference) to the simulator.

Note that we have received various designed trajectories (references) in the trajectory design process from the problems with different dynamic state error weights.

Likewise, we will receive as many as the resulting actual tracking trajectories since each is the tracking result of the corresponding designed trajectory. We map these actual tracking results in a 'pseudo-Pareto frontier'.

A mapping 'pseudo-Pareto frontier' is a plot providing the information of the actual tracking trajectories. It is generated with the following procedure. Once we receive a resulting actual tracking trajectory, we record the actual dynamic state error and the actual acceleration. Using these two pieces of information, we plot a dot on a cost-error plane where the x-axis and y-axis represent the cost function and the dynamic state error, respectively.

A collection of the dots will be received once we receive different resulting actual tracking trajectories. These dots in the cost-error figure shows the similarity pattern with a typical Pareto frontier. Thus, we name this figure pseudo-Pareto frontier. Notice that every dot on the pseudo-Pareto frontier matches the corresponding dot in the Pareto frontier for trajectory design.

## IV. SOLVER SETUP

Our solver is built based on approximations and equivalent transformations in direct collocation and the dynamic state estimator. This section details these mathematical bases.

We solve the optimal trajectory optimization problem using direct collocation method in this research. The first step in a direct method is to discretize the dynamics, the objective functions, and the dynamic state error estimation (Equation (24)-(25)). After the discretization, the optimization problem is transferred to a nonlinear programming problem. A MATLAB built-in function (fmincon) is then utilized for solving this NLP problem before further final modification, e.g., smoothen.

In the discretization of the dynamic constraints, Equation (21), we evenly divide the entire time horizon into 60 timepieces. Thus, each time segment lasts 1/60 second. We only take the knot points into calculation in trajectory optimization.

The dynamic constraint, the differential equation in Equation (26), is approximated at knot points to maintain its property during discretizing. Equation (27) specifies the principle of the trapezoid in the dynamic constraints' discretization [19].

$$\dot{x} = f(t) \tag{26}$$

$$x(t_k) - x(t_{k-1}) = \frac{1}{2} \cdot \Delta t \cdot [f(t_k) + f(t_{k-1})] \tag{27}$$

Here, $\Delta t$ is the time segment and is equal to the difference between the $t_k$ and $t_{k-1}$. As mentioned before, it is 1/60 second in our solver.

As for the dynamic state error estimation, Equation (24)–(25), several deductions are made before further actual application.

First, the control constant, $p$ in Equation (24), is determined by the controller defined in Equation (12) - (14).

$$p = 1 - e^{\lambda \cdot \frac{t}{n}} \tag{28}$$

The $\lambda$ in Equation (28) is the equivalent first-order eigenvalue. For example, the controller with closed-loop system eigenvalue pair [-20 -200] is approximated with a single first-order eigenvalue, $\lambda = -20$.

Substituting Equation (28) into Equation (24)–(25), we receive the dynamic state error estimation in the form of integration in Equation (29).

$$e(t) = e^{\lambda \cdot t} \cdot \int_0^t v_{ref}(t) \cdot e^{-\lambda \cdot t} dt \tag{29}$$

Alternatively, Equation (29) can also be obtained by solving an equivalent differential equation directly. However, it can lead to an unreasonable dynamic state error estimation for a discrete system other than Equation (24).

For example, we may discretize Equation (29) into Equation (30). However, Equation (30) is not identical to Equation (24) unless n is infinite.

$$e(t,n) = e^{\lambda \cdot t} \cdot \sum_{i=1}^{n} v_{ref}\left(\frac{t}{n} \cdot i\right) \cdot e^{-\lambda \cdot \left(\frac{t}{n} \cdot i\right)} \cdot \frac{t}{n} \tag{30}$$

Deciding on the proper dynamic state error estimation among Equation (24), (29), and (30) can be controversial. The basic idea is that Equation (29) is suggested for a continuous system. On the contrary, Equation (24) is encouraged for a discrete system or a discrete controller. True, the controller and the system are both continuous in this problem. However, the nature of direct collocation makes them discretized in the trajectory optimization process.

In this research, we utilize Equation (29) to estimate our dynamic state error. A further trade-off between these two formulas is beyond our scope in this research and is still an open question.

Nevertheless, we still cannot calculate the integration in Equation (29) since our calculation is based on the data on knot points. To solve this problem, the integration is approximated by Equation (31).

$$\int_0^{N \cdot \Delta t} h(t) dt = \begin{bmatrix} \frac{1}{2} & 1 & 1 & \cdots & 1 & 1 & \frac{1}{2} \end{bmatrix} \cdot \begin{bmatrix} h(0) \\ h(\Delta t) \\ h(2 \cdot \Delta t) \\ \vdots \\ h((N-2) \cdot \Delta t) \\ h((N-1) \cdot \Delta t) \\ h(N \cdot \Delta t) \end{bmatrix} \cdot \Delta t \tag{31}$$

## V. RESULTS

The results for the optimal trajectory design for the eigenvalue pair [-20, -200] is demonstrated in Figure 4.

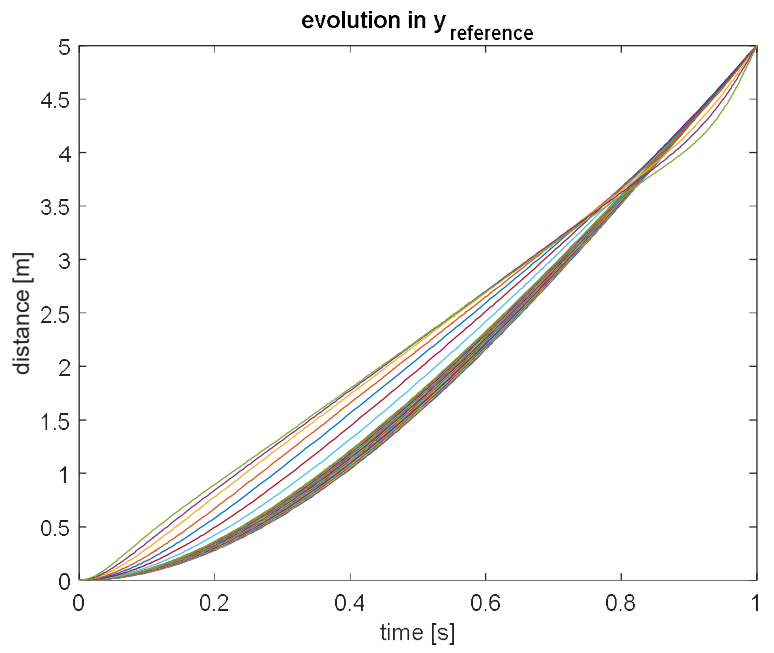

Fig. 4. The result in trajectory design process with different coefficient $\mu$

With the increase of the coefficient $\mu$ in Equation (23), the resulting trajectory receives an evolution. On the other hand, the actual tracking resulting trajectory also updated (Figure 5) for the eigenvalue pair [-20, -200].

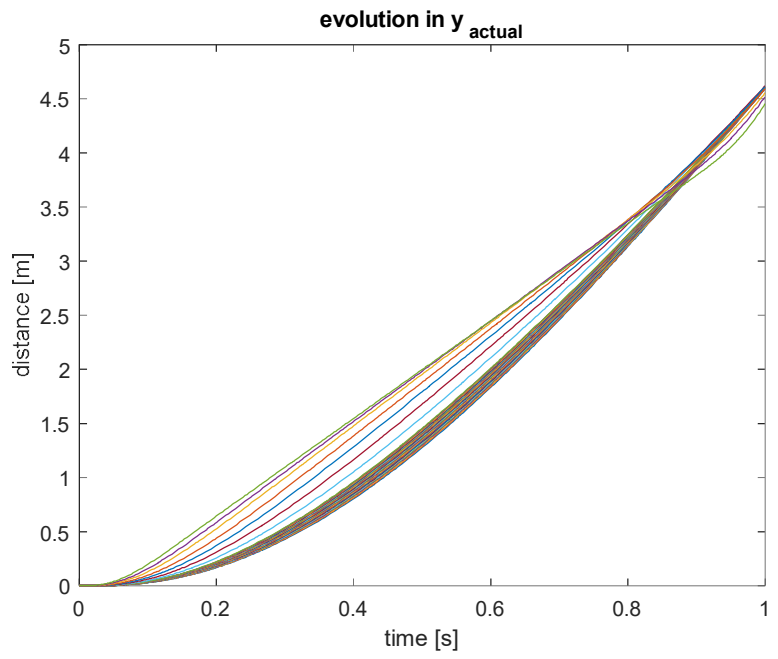

Fig. 5. The actual resultant trajectories in tracking control process

The blue curve in Figure 6 shows the Pareto frontier of the trajectory design for the eigenvalue pair [-20 -200]. The $x$ axis represents the cost function defined in Equation (17). The $y$ axis represents the integral of the dynamic state error squared Equation (32).

$$e(t) = \int_0^1 e(t)^2 dt \tag{32}$$

The blue curve shows a trade-off between the trajectory plan (cost function) and the controller design (dynamic state error).

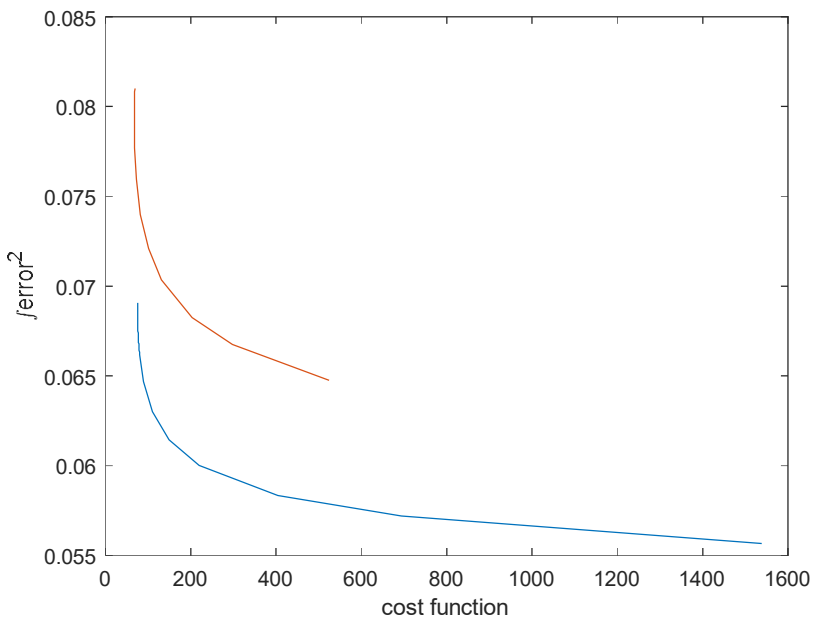

Fig. 6. The blue curve is the Pareto frontier of the trajectories designed in the trajectory design process. The red curve is the 'pseudo Pareto frontier' of the actual trajectories in the tracking control process.

Note that each data on the blue curve in Figure 6 corresponds to a resultant tracking trajectory in Figure 4. We know the exact value of the cost function as well as the dynamic state from each tracking result in Figure 4. Thus, we can plot a data dot in Figure 6 for each actual tracking trajectory in Figure 4. As a result, we receive the red curve in Figure 6.

It is also worth mentioning that each dot on the trajectory Pareto frontier on the blue curve in Figure 6 corresponds to a dot of its tracking result dot on the red curve in Figure 6. We call this relationship 'mapping'. The Pareto frontier (blue) is mapped to a 'pseudo Pareto frontier' (red) which represents the actual result in tracking process.

After zooming the initial part of the pseudo Pareto frontier, we can see an atypical shape in Pareto frontier (Figure 7). Both the cost and dynamic state error decrease near the neck of the 'pseudo Pareto frontier'.

This near-neck atypical shape in 'pseudo Pareto frontier' also occurs in this system equipped with the rest of our controllers with different eigenvalues.

Our ultimate goal is to find the actual moving trajectory with the least cost function. Thus, the best compromised choice lies at the dot with the least cost in pseudo Pareto frontier (red curve in Figure 6 or Figure 7)

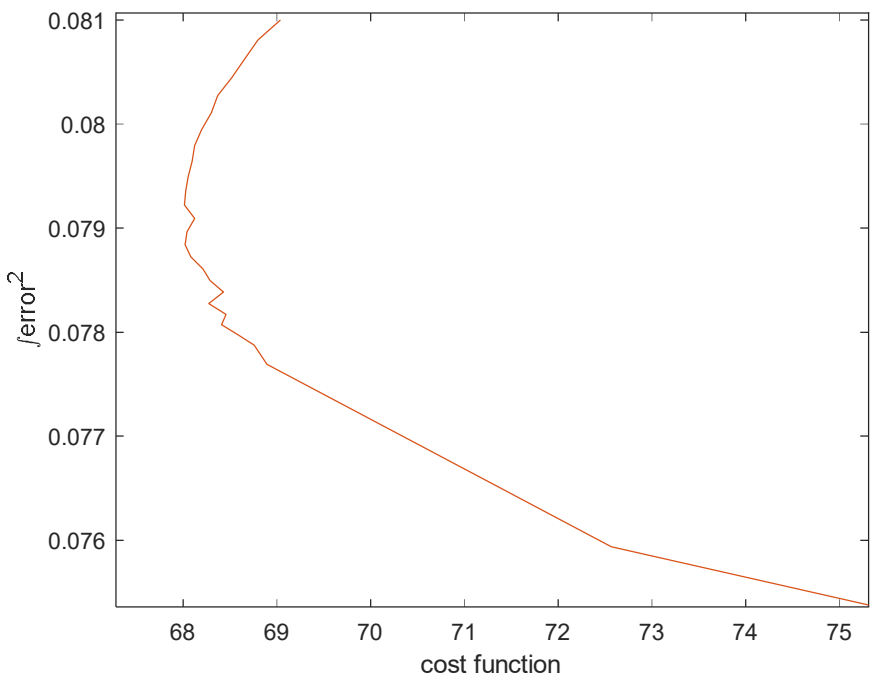

Fig. 7. The 'neck' part of the 'pseudo Pareto frontier' zoomed from Figure 6.

Notice that we do not receive the actual tracking result in the trajectory design process. So, the corresponding data dot mapping to the dot with the least cost in pseudo Pareto frontier is found on the Pareto frontier in the Figure 6 blue curve. The dot is indicated in Figure 8.

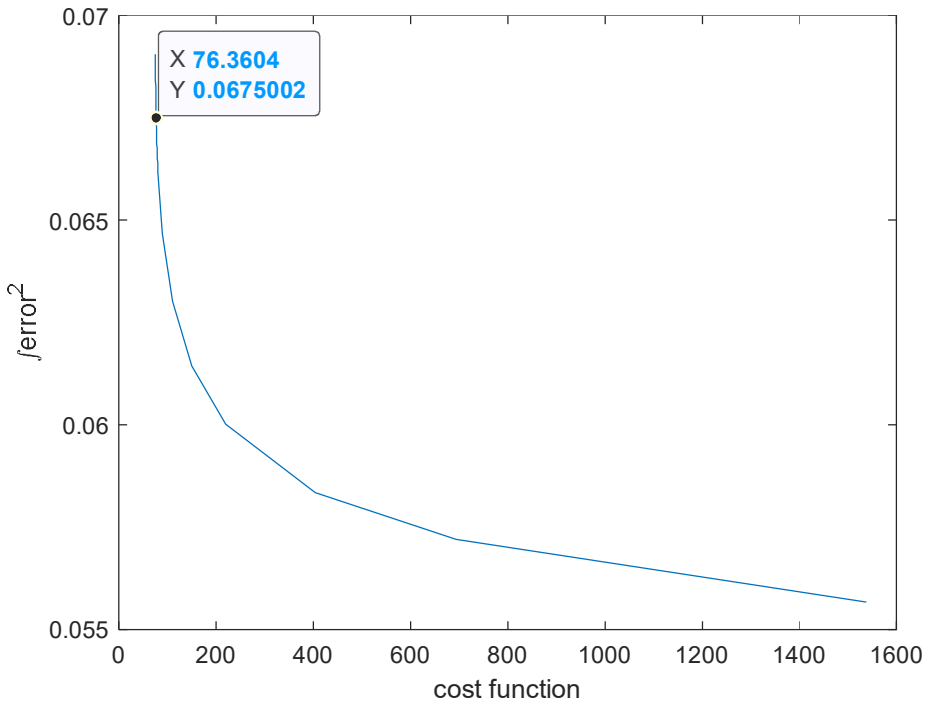

Fig. 8. The best compromised choice in the Pareto frontier in the trajectories design process.

The similar results can be found with different LQR controller settings (eigenvalues). The relationship between the 'pseudo Pareto frontier' and the controller is detailed in the next Section.

## VI. STIFFNESS OF THE CONTROLLER

The Pareto frontiers (blue) and the mapped pseudo Pareto frontiers (red) for the UAV with the rest LQR parameter settings (Eigenvalue pairs) are plotted in Figure 9-11. They are the results from the LQR eigenvalue pairs [-10, -100], [-30, -300], and [-50, -500], respectively.

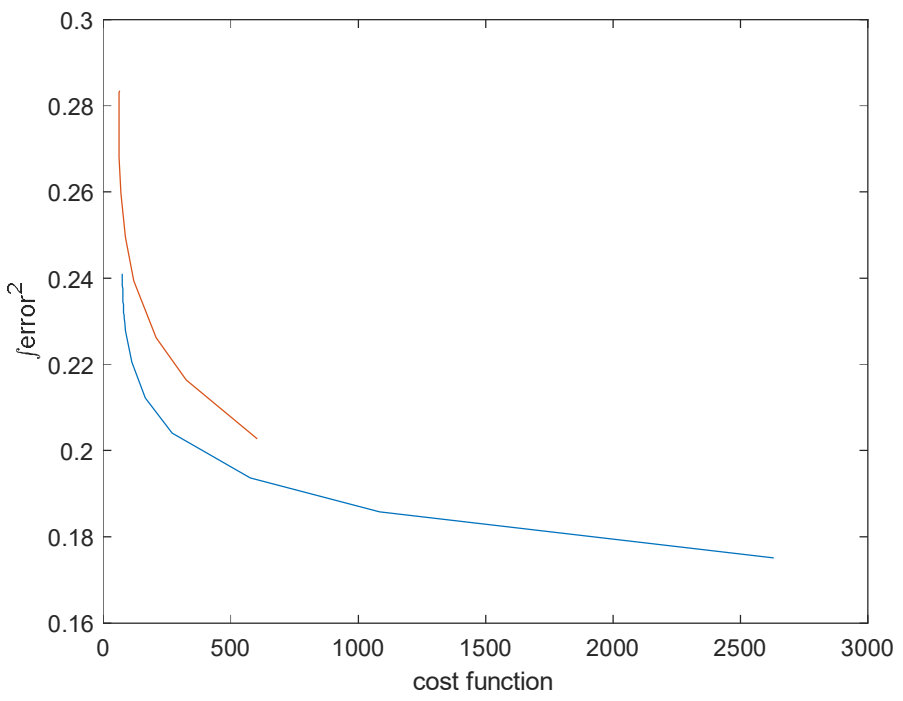

Fig. 9. Pareto frontier (eigenvalue pair [-10, -100])

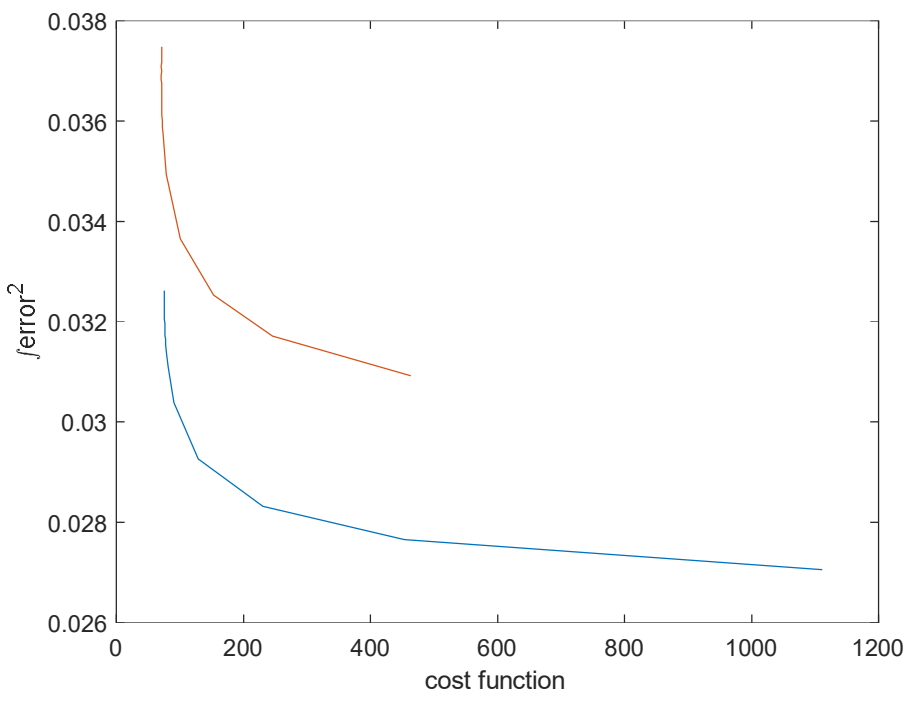

Fig. 10. Pareto frontier (eigenvalue pair [-30, -300])

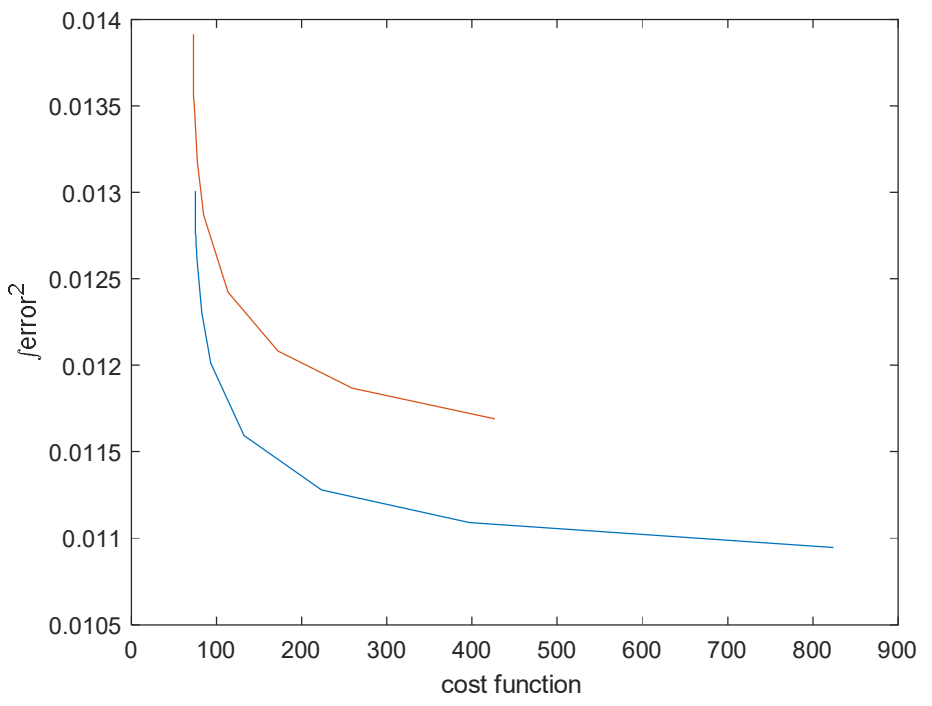

Fig. 11. Pareto frontier (eigenvalue pair [-50, -500])

All the blue curves in the trajectory in Figure 6, 9-11 are typical Pareto frontiers in a multiple-object optimization (trajectory design problem). While the mapped red Pareto frontiers represent the cost and the error in tracking results.

In addition, each of these pseudo Pareto frontiers contains the atypical shape demonstrated in Figure 7. We find the data dot receiving the lowest cost in the pseudo Pareto frontier and find the corresponding data dot in the Pareto frontier as what we were doing in Figure 8.

The relevant corresponding data dot in the Pareto frontiers with the rest LQR parameter settings (Eigenvalue pairs) are marked in Figure 12-14 (eigenvalue pairs [-10, -100], [-30, -300], and [-50, -500], respectively).

These dots in the Pareto frontiers match the trajectory receiving the lowest cost in its psuedo Pareto frontiers. All of them are near the neck of the Pareto frontiers.

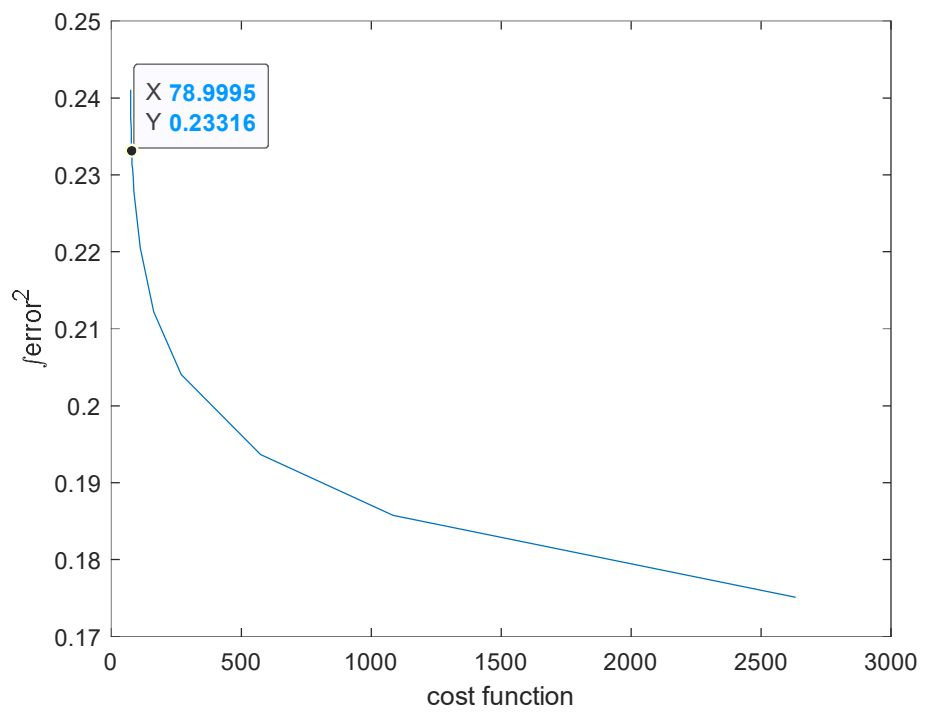

Fig. 12. The corresponding data dot in the Pareto frontier (eigenvalue pair [-10, -100])

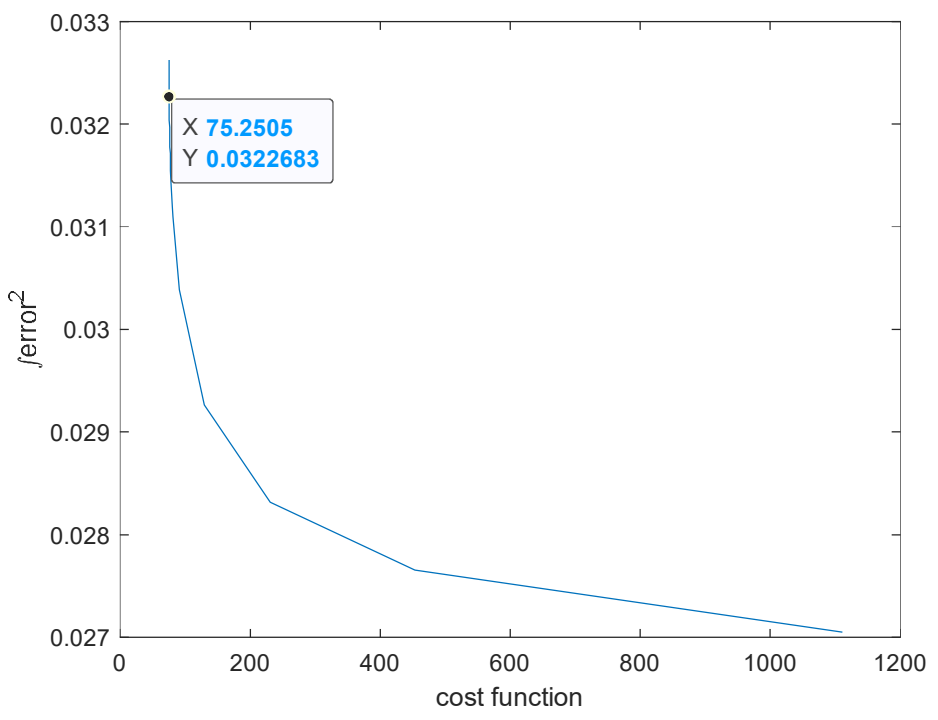

Fig. 13. The corresponding data dot in the Pareto frontier (eigenvalue pair [-30, -300])

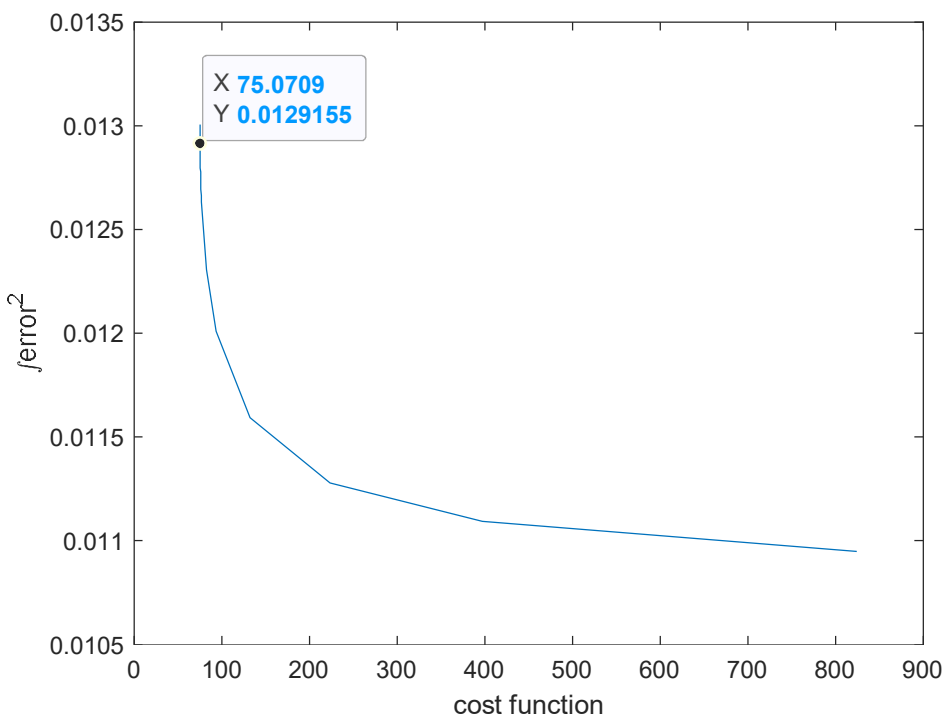

Fig. 14. The corresponding data dot in the Pareto frontier (eigenvalue pair [-50, -500])

We model the atypical shape in pseudo Pareto frontier, e.g., Figure 7, with a physical model which is detailed in the Figure 15-16.

A rope satisfies the Hooke's law, Equation (33), and is fixed on the wall on both sides.

$$F = -k \cdot {}_{\triangle}x \tag{33}$$

We exert 1 Newtown force in the middle of the rope (Figure 15). The consequence is that the rope extends, leaving from the initial level position (Figure 16).

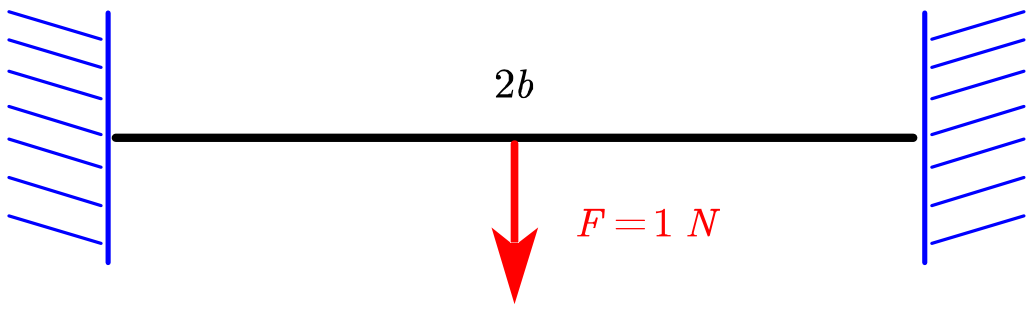

Fig. 15. Exert 1 Newtown force on a rope (length: $2b$) fixed on the wall on both sides.

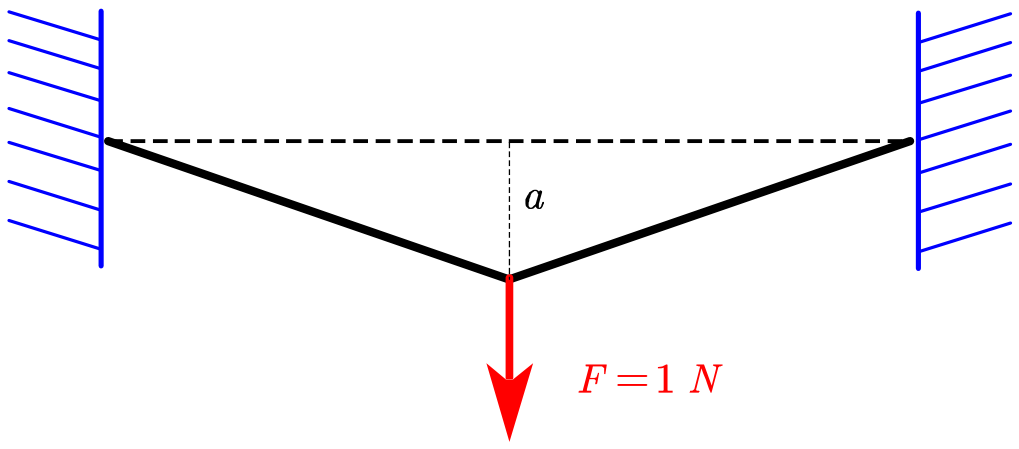

Fig. 16. The rope obeys the Hooke's law and changes its shape.

We assume that the initial length of the rope is $2b$. Also, we assume that the middle of the line displace $a$ along the direction of the force.

Based on these assumptions, the spring constant, $k$ in equation (33), is calculated in Equation (34).

$$k = \frac{1}{4 \cdot a \cdot \left\{1 - \left[1 + \left(\frac{a}{b}\right)^2\right]^{-\frac{1}{2}}\right\}} \tag{34}$$

As indicated in Figure 17, we take the largest decrease from initial cost in the cost function for the parameter $a$ in Euqation (34). While we take the initial cost in Figure 17 $2b$ in Equation (34).

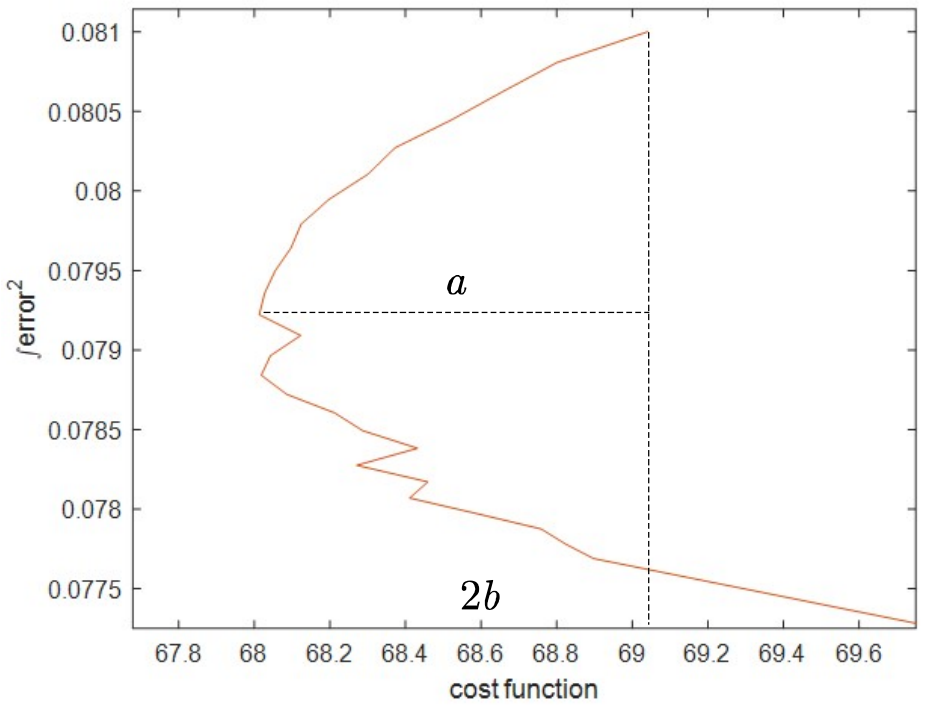

Fig. 17. $2b$ is the initial cost. $a$ is the largest decrease from initial cost in the cost function. (Eigenvalues: [-20 -200])

Consequently, we can calculate a spring constant based on Equation (34) for the eigenvalue pairs [-20 -200]. The spring constants for the rest eigenvalue pairs, namely [-10, -100], [-30, -300], and [-50, -500], can also be obtained.

The spring constants for its corresponding eigenvalues are plotted in Figure 18.

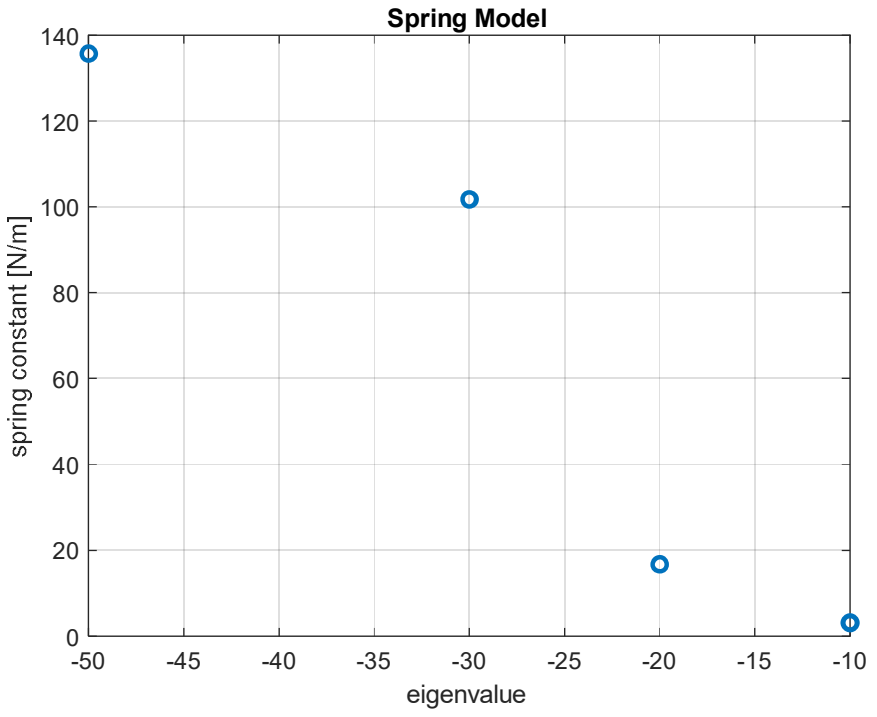

Fig. 18. The spring model for the pseudo Pareto frontiers.

The conclusions and discussions are made in the next section.

## VII. CONCLUSIONS AND DISCUSSIONS

When the LQR controller is more aggressive (e.g., a larger equivalent eigenvalue), the following results are observed:

**Result 1.** The designed cost function becomes closer to the real cost function.

This can be found in 6, 9-11; the designed cost function (blue curve) is closer to the true tracking result (red curve) with a larger eigenvalue.

It indicates that the aggressive controller makes the tracking result closer to the trajectory we designed.

**Result 2.** Best compromised result with the lowest actual cost becomes closer to the head region of the Pareto Frontier of trajectory design.

The data dots illustrated in 8, 12-14 are the best compromised result which correspond to the actual tracking result receiving the lowest cost in the pseudo Pareto frontier. These data points are closer to the head of the Pareto Frontier of trajectory design when the LQR equivalent eigenvalue is larger.

The actual tracking result receiving the lowest cost does not happen for the case when $\mu = 0$, which corresponds to the initial (top) data dot in the Pareto frontier. Instead, a better checking is rewarded by sacrificing the cost in the trajectory trajectory plan process. Though the trajectory designed is endowed with the lowest cost when $\mu = 0$, the tracking process introduces the dynamic state error, making the actual cost not the lowest compared with the compromised trajectory possessing better tracking properties.

An aggressive controller weakens the effect of this compromise; the nature of these strong controllers provides the less dynamic state error. That's why the best compromised trajectory receiving the lowest cost in the tracking process locates closer to the initial data dot (head region) in the Pareto frontier when the controller is more aggressive.

**Result 3.** The equivalent spring model becomes more stubborn.

The spring coefficient ($k$) in the physical model in Figure 15-16 represents the stubbornness of the rope. The rope with a larger $k$ is more stubborn, requiring more force to change its shape.

Likewise, a large spring coefficient ($k$) for our pseudo Pareto frontier, e.g., Figure 17-18, indicates that the corresponding pseudo Pareto frontier is more stubborn; the atypical shape in the pseudo Pareto frontier shown in Figure 17 becomes less obvious.

The nature of the LQR parameters contributes to this consequence. Since the aggressive controller lets the actual moving trajectory better obey the designed trajectory as well as the dynamic state error, the pseudo Pareto frontier is closer to its corresponding Pareto frontier. However, there is no atypical shape in the Pareto frontier (this special shape only exists in the pseudo Pareto frontier). Thus, an aggressive controller eliminates this atypical shape in a larger degree. That is the underlying reason for receiving Figure 18.

The pseudo Pareto frontier in this paper presents various advantages in analyzing the dynamic state error. The method of analyzing the controller in the view of a physical model can be extended to other controllers.

Further steps can be the selection of the formula utilized in estimating the dynamic state error for a cost-error scenario.